\documentclass[conference]{IEEEtran}
\makeatletter
\IEEEoverridecommandlockouts
\usepackage{cite}
\usepackage{amsmath,amssymb,amsfonts}
\usepackage{graphicx}
\usepackage{xcolor,soul}
\usepackage[flushleft]{threeparttable}
\usepackage{booktabs}
\def\BibTeX{{\rm B\kern-.05em{\sc i\kern-.025em b}\kern-.08em
	T\kern-.1667em\lower.7ex\hbox{E}\kern-.125emX}}
\usepackage{algorithm}
\usepackage{algpseudocode}
\usepackage{bm}
\usepackage{array}
\usepackage{multirow}
\usepackage[draft,footnote,nomargin]{fixme}

\begin{document}
\IEEEoverridecommandlockouts
\title{
Virtual Reality Gaming on the Cloud: \\A Reality Check

	\thanks{This work is supported in part by the Ontario Center of Innovation (OCI) ENCQOR 5G development program and the Natural Sciences and Engineering Research Council of Canada (NSERC) under Grant RGPIN-2020-04661.}
	\thanks{(\textit{Corresponding author: Xiao-Ping Zhang}, e-mail: xzhang@ryerson.ca)}
}

\author{\IEEEauthorblockN{Sihao Zhao$^\ddag$, Hatem Abou-zeid$^*$, Ramy Atawia$^*$,\\ Yoga Suhas Kuruba Manjunath$^\ddag$, Akram Bin Sediq$^*$, Xiao-Ping Zhang$^\ddag$
	}
	\IEEEauthorblockA{$\ddag$Department of Electrical, Computer and Biomedical Engineering, Ryerson University, Toronto, ON, Canada \\
		*Ericsson Canada, Ottawa, Canada}\\
}

\maketitle

\begin{abstract}
Cloud virtual reality (VR) gaming traffic characteristics such as frame size, inter-arrival time, and latency need to be carefully studied as a first step toward scalable VR cloud service provisioning. To this end, in  this  paper we  analyze  the  behavior  of  VR gaming traffic  and  Quality of Service (QoS)  when  VR  rendering  is  conducted remotely  in  the  cloud. 
We first build a VR testbed utilizing a cloud server, a commercial VR headset, and an off-the-shelf WiFi router. 
Using this testbed, we collect and process cloud VR gaming traffic data from different games under a number of network conditions and fixed and adaptive video encoding schemes.
To analyze the application-level characteristics such as video frame size, frame inter-arrival time, frame loss and frame latency, we develop an interval threshold based identification method for video frames. 
 Based on the frame identification results, we present two statistical models that capture the behaviour of the VR gaming video traffic. The models can be used by researchers and practitioners to generate VR traffic models for simulations and experiments - and are paramount in designing advanced radio resource management (RRM) and network optimization for cloud VR gaming services.
To the best of the authors' knowledge, this is the first measurement study and analysis conducted using a commercial cloud VR gaming platform, and under both fixed and adaptive bitrate streaming. We make our VR traffic data-sets publicly available for further research by the community.

\end{abstract}



\section{Introduction}
Virtual reality (VR) is becoming popular in a variety of applications such as gaming, healthcare and education \cite{elbamby2018toward,lai2019furion,sun2019communications}. In the VR gaming sector, the user needs a head mounted display (HMD) device or headset and controllers to build an immersive and interactive gaming environment. VR games require intensive computational resources to render and display the video frames in a short motion-to-photon time so that the players can obtain a real-time immersive experience. This is challenging for wearable headsets, that are constrained by size, weight and power consumption. As an alternative solution, headset products such as Valve Index \cite{valveindex}, Oculus Rift \cite{oculushmd}, HTC Vive \cite{htcvive}, and Sony PlayStation VR \cite{sonystationvr} enable a gaming PC with dedicated graphic processing units (GPUs) to run the VR games and stream the rendered videos to the headset via cable or wireless connection. More recently, the latest product of Oculus Quest 2 HMD offers the capability of both standalone game running and remote VR rendering via wireless connectivity to the network \cite{oculushmd}.

By offloading the rendering task to a gaming PC, the burden of the headset is relieved. However, a higher data transfer rate is required to transmit the VR video frames, which are much larger than typical videos, to the headset. This data transfer link can be the Internet, local cable network, WiFi and USB, among others.
In the case when the gaming PC is connected to the headset via the Internet, it is referred to as cloud VR gaming. Nowadays, there are a number of cloud VR gaming service providers such as Shadow \cite{shadow}, Maximum Settings \cite{maximunsettings} and Paperspace \cite{paperspace}. They provide dedicated gaming PCs with several hundreds of mega bit per second Internet access.

In the existing literature of the cloud gaming testbeds, Xue et al. investigated the performance of the CloudUnion gaming service and found that there was large queuing delay, conservative video rate recommendations were adopted, and the difference between UDP and TCP data transmission \cite{xue2014playing}. Huang et al. developed a cloud gaming platform, which offered shorter frame processing delay compared with other cloud gaming services \cite{huang2013gaminganywhere}. Chen et al. measured the latency of two cloud gaming services and found over 100 ms delay for both of them \cite{chen2011measuring}. These studies only consider the ordinary PC games rather than VR games.

There are few recent studies on the VR gaming traffic. Li et al. build a local VR gaming testbed based on the ALVR software to evaluate the impact of latency, bandwidth and packet loss on the the quality of service (QoS) and quality of experience (QoE) \cite{li2020performance}. The study in \cite{liu2018cutting} aims to reduce the latency in a wireless VR system, and a testbed is developed to evaluate the end-to-end latency and frame loss. However, they only emulate the cloud gaming scheme by using local work stations.

In this paper, we aim to understand the behavior of VR gaming traffic and QoS when VR rendering is conducted remotely in the cloud. To this end, 
we have implemented a testbed using a \textit{commercial} cloud gaming server \cite{paperspace},
the Oculus Quest 2 (one of the latest VR headsets) \cite{oculushmd}, and an off-the-shelf 5GHz WiFi router.
 Using this testbed, we collect and process cloud VR gaming traffic data from different games under a number of network conditions and fixed and adaptive video encoding schemes. We also capture and identify all the traffic flows involved in the VR gaming application including video, audio, and control data flows. To analyze the application-level characteristics such as video frame size, frame inter-arrival time, frame loss and frame latency, we develop an interval threshold based identification method for video frames. We find that adaptive bit rate encoding provides a more stable gaming performance under stricter bandwidth throttling than fixed bit rates.
 Based on the frame identification results, we present two statistical models that capture the behaviour of the VR gaming video traffic. The models can be used by researchers and practitioners to generate VR traffic models for simulations and experiments - and are helpful in designing advanced radio resource management (RRM) and network optimization for cloud VR gaming services.
In summary the contributions of this paper are:
\begin{itemize}
    \item We measure and analyze the impact of limited bandwidth on VR game video frame metrics and latency using a commercial cloud gaming server under different video encoding and bit rate settings.
    \item We show that the VR video frame size and inter-arrival time can be modeled as loglogistic and Burr distributions, respectively. These mathematical representations of VR frames have not been presented in literature using traces collected from a commercial cloud testbed.
    \item We develop an interval threshold-based method to identify the application-level video frame characteristics using only the raw data packets. This avoids the need for packet decoding or deep packet inspection.
    \item Our VR traffic data and performance measurements under different data rate throttling settings at the WiFi router, with both fixed and adaptive video bit rate transmission schemes, are publicly available for further analysis by the research community\cite{VRdataset}.
\end{itemize}


\section{Testbed Implementation}
\subsection{Cloud Gaming Testbed}
As shown in the top left in Fig. \ref{fig:cloudscheme}, we use the Paperspace cloud PC to run the VR games. The cloud PC is equipped with Intel Xeon E5-2623 CPU with 30 GB RAM, and Nvidia P4000 GPU with 8 GB memory, which is sufficient for VR game rendering. The Internet access bandwidth is 120 Mbps. The gateway device is a 5 GHz band WiFi router, which offers up to 866 Mbps wireless bandwidth. The Internet is connected to the router through cable. The VR headset used is the latest Oculus Quest 2 with 256 GB storage space and connected directly to the WiFi gateway through 5 GHz frequency band.

We install Virtual Desktop software \cite{vrdesktop} on the cloud PC. The client is installed at the headset to transmit and receive VR gaming data. Wireshark \cite{wireshark} is installed on the cloud server to capture all the communication flows to and from the server.

\subsection{Local Gaming Setup}
\begin{figure}
	\centering
	\includegraphics[trim=70 0 100 -10, width=0.6\linewidth]{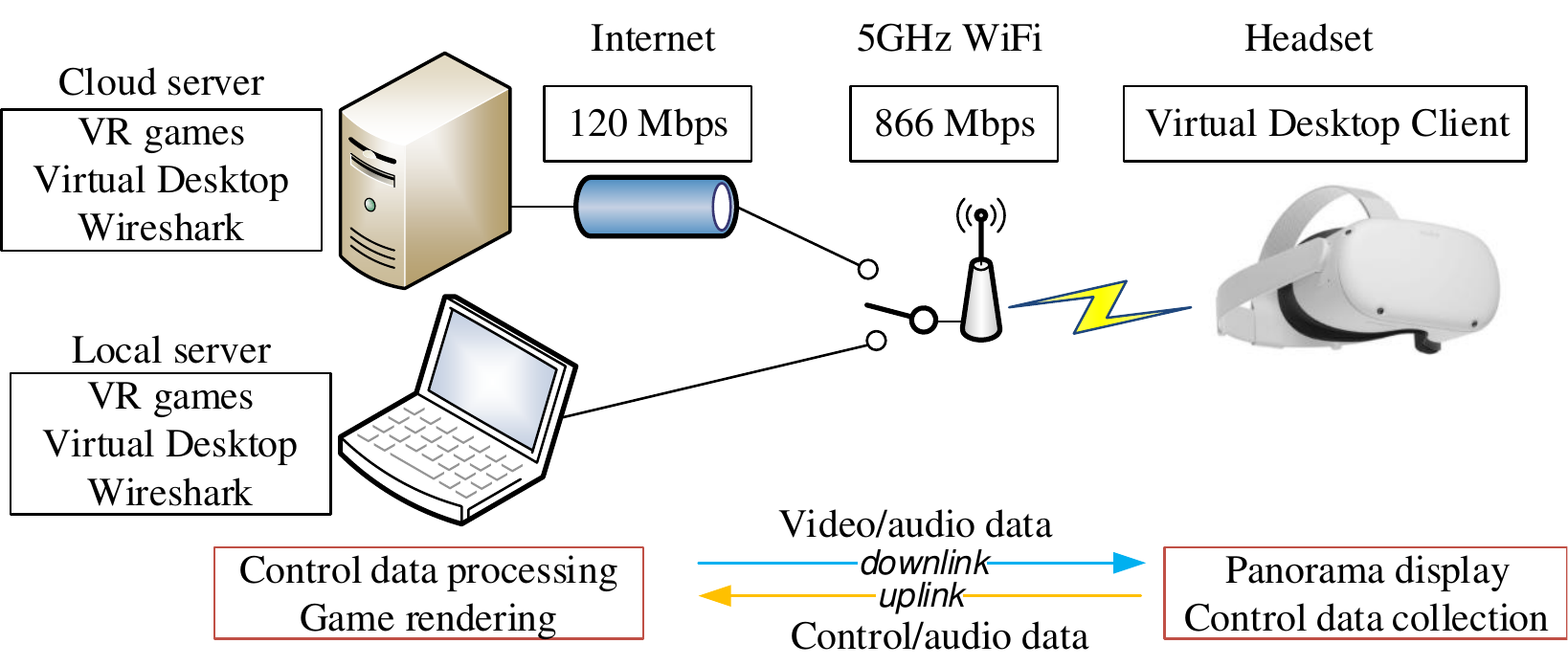}
	\vspace{-0.2cm}
	\caption{Cloud and local gaming diagram
	}
	\label{fig:cloudscheme}	
	\vspace{-0.5cm}
\end{figure}

We also setup a local VR gaming system to collect traffic and latency measurements for performance comparison to the cloud setup. This is illustrated in Fig. \ref{fig:cloudscheme} with the local server connection. A laptop with i7 9750H CPU, 32GB RAM, and Nvidia Quadro T1000 GPU with 4GB memory is used to run the VR games locally. The laptop is connected to the LAN port on the WiFi router by cable and supports 1000 Mbps bandwidth.

As in the cloud setup, we also install Virtual Desktop software on the local laptop to work with the client on the headset for gaming data streaming. We use Wireshark on the local laptop to capture the gaming traffic.

\subsection{VR Games for Test}
We use two VR games, namely Beat Saber and Steam VR Home for test. Beat Saber is a sports game with real-time control and fast moving objects in the field of view. The Steam VR Home is a relatively stationary room scene with a few moving objects such as birds and clouds. 

\subsection{Streaming Settings}
The main parameters set in the Virtual Desktop software are ``90 Hz'' frame rate, ``High'' graphics quality, and ``40 Mbps'' streaming bit rate limit. Enabling the adaptive video bit rate setting will result in a varying rate below the maximum based on the available bandwidth and latency. On the other hand, disabling the adaptive video bit rate results in a fixed streaming rate at the maximum value of 40 Mbps. We use Wireshark to record about 30 s of traffic data from the two VR games with both fixed and adaptive video bit rate setting, in both the local and cloud gaming cases. 

\section{VR Gaming Traffic Flows}
We have identified three port pairs 
used by Virtual Desktop between the cloud/local server and the VR headset. The first port pair is used to transfer video frame data, and has the largest data volume. The other two port pairs are used for audio data transfer between the server and the headset, and for control data flow from the headset to the server as well as their acknowledgments. We will analyze these three data flows with a focus on the video flow.

\subsection{Video Flow}\label{videotraffic}
We observe that most of the packets in both the cloud and local streaming cases have a length of 1514 bytes. There is a series of the 1514-byte packets followed by a smaller packet. All these packets in such a group are transmitted at the same instant. There are usually two groups of these packets that are sent successively. We infer that these two groups are for the views on left and right eyes, and they belong to the same frame. Since H.264 or HEVC encoding is used, we also infer that the seldom groups with larger total sizes are I or P-frames and the other smaller groups are B-frames. The probability density function (PDF) of the downlink or server-to-headset video packet transmission interval for the VR game Beat Saber are shown in Fig. \ref{fig:pktinterPDF}.




The data traffic from the headset to the server or uplink contains only ACK packets. For both cloud and local cases, the packet size remains constant at 60 byte, and the data rate is lower than 0.1 Mbps.

\subsection{Control Flow}
In the uplink of this data flow, the headset sends control data from both the headset and the controllers to the server. In the downlink, the server sends ACK packets to the headset. The uplink packets have a constant size of 358 byte and a data rate smaller than 0.4 Mbps. The downlink packets are ACK packets with fixed size of 54 byte.



\subsection{Audio Flow}
In this flow, there are audio data packets in both the uplink and downlink. The server sends 1222-byte audio packets to the headset in the downlink and the headset sends 60-byte ACKs in the uplink. The headset sends 390-byte packets to the server and the server sends 54-byte ACKs in the downlink. The data rates of the uplink and the downlink are about 0.33 Mbps and 1.57 Mbps, respectively.



\section{VR Video Frame Identification Method}
Among the three VR gaming flows, we are most interested in the video traffic stream. It has a large data rate and volume and affects the user experience the most. In this section, we propose a method to identify the video frame size and inter-arrival time. Then, based on the identified frames, we further compute the video frame loss and frame latency.

\subsection{Video Frame Size and Inter-arrival Time Identification}\label{fsitmethod}
Without decoding the traffic data, we are not able to obtain knowledge on the application-level information such as VR frame sizes and inter-arrival time of the frames. However, by observing the traffic, we infer that the data packets within one frame are temporally close. We have observed that there are several 1514-byte packets along with a smaller packet transferred at the same time, and there are  two such successive sequences, which are temporally close to each other. We therefore infer that these two successive sequences of packets form a VR video frame. For both the cloud and local gaming cases, the traffic patterns are found to be similar.

\begin{figure}
\centering
	\includegraphics[trim=70 -20 70 0 ,width=0.7\linewidth]{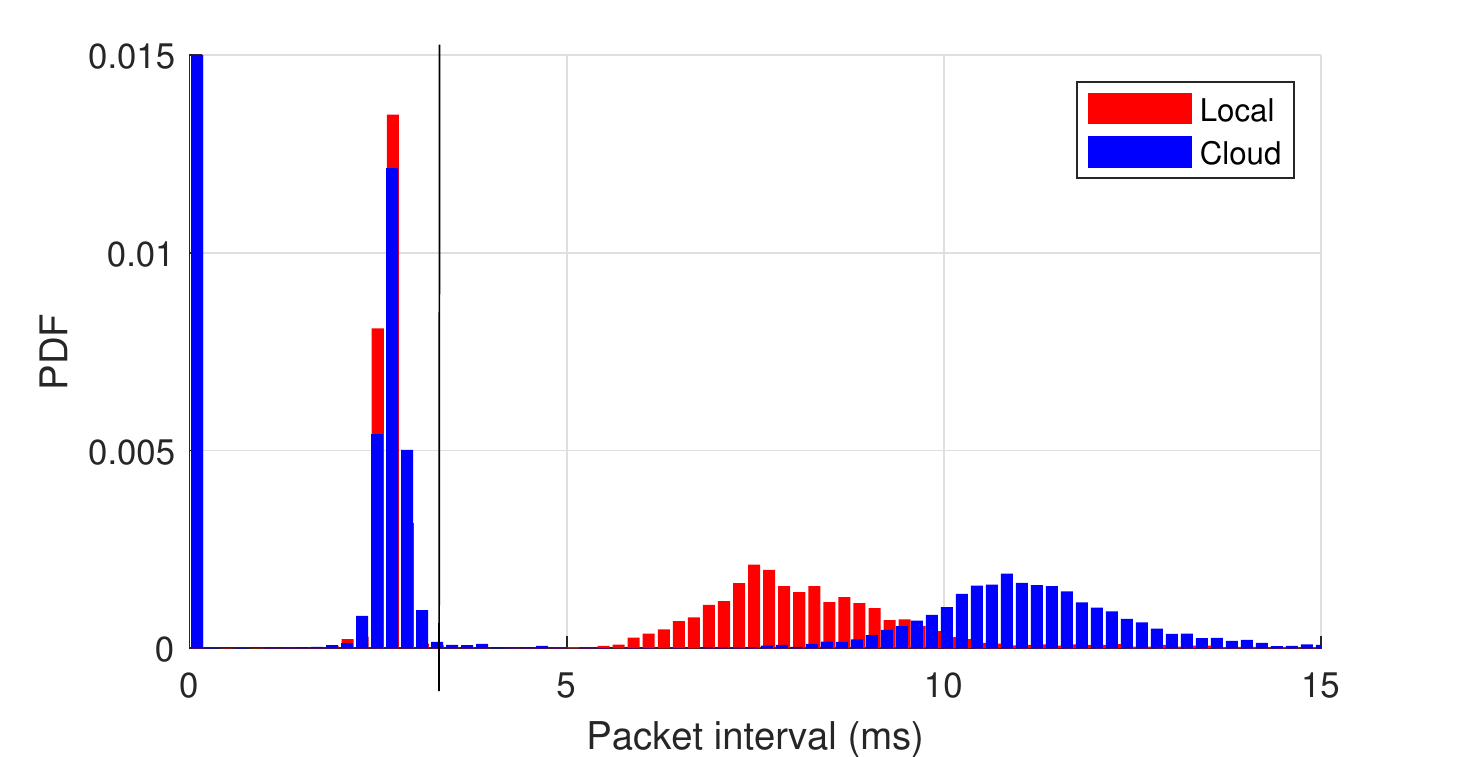}
	\vspace{-0.6cm}
	\caption{PDF of video data packet transmission interval for local and cloud streaming (Beat Saber). The actual peak value at 0 ms is over 0.9, but is limited by the figure scale.
}
	\label{fig:pktinterPDF}
	\vspace{-0.4cm}
\end{figure}

As can be seen from the histogram of the packet interval in Fig. \ref{fig:pktinterPDF}, a lot of packets in both cloud and local cases have zero interval. These are the data packets transmitted at the same instant. In the same figure, the second largest peak from the left represents the interval between two packet sequences with several 1514-byte packets followed by a smaller packet. The other two lower peaks on the right represent the intervals between successive video frames. The left two higher peaks and the right two lower peaks represent inner and inter frame intervals, respectively. They have a clear gap between 3 and 5 ms. The other game show similar patterns. This will help us to identify the video frames from the raw packets.

Based on the above observation and analysis, we use an interval threshold, denoted by $\Delta t_{thr}$ to cluster the packets that are close to each other into one ``video frame''. Specifically, we treat the successive packets that have an interval shorter than $\Delta t_{thr}=3$ ms as within the same video frame. In this way, we are able to identify each frame and estimate the frame size. The frame inter-arrival time is estimated by subtracting the transmission times of the very first two data packets in two successive frames.


\subsection{Frame Loss Rate}\label{lossmethod}
Note that the traffic data is collected at the server side. Normally, the server receives ACK packets if the headset receives one or more video packets. If the server does not receive any ACK, we know that the video frame is not transferred successfully.
Based on the identification of frames, we try to find the ACK packets for every identified video frame. Particularly, if there is no ACK for a certain frame, then we treat this frame as lost. Therefore, the frame loss rate is expressed as
\begin{align}
	\text{Frame Loss Rate}=\frac{\text{Number of Frames without ACK}}{\text{Total Number of Frames}}.\nonumber
\end{align}


\subsection{Frame Latency} \label{fptmethod}
Based on the frame identification, we compute the interval between the transmission of the first data packet and the reception of the last ACK packet in one frame, i.e., 
\begin{align}
	&\text{Frame Latency}=\nonumber\\
	&(\text{RX Time of last ACK})-(\text{TX Time of 1\textsuperscript{st} data packet}).\nonumber
\end{align}

\section{VR Video Frame Characterization with Fixed Bit Rate}\label{fixedrate}
\subsection{Metrics without Data Rate Limit}
\begin{figure}
	\centering
	\includegraphics[trim=70 -20 40 -20 ,width=0.7\linewidth]{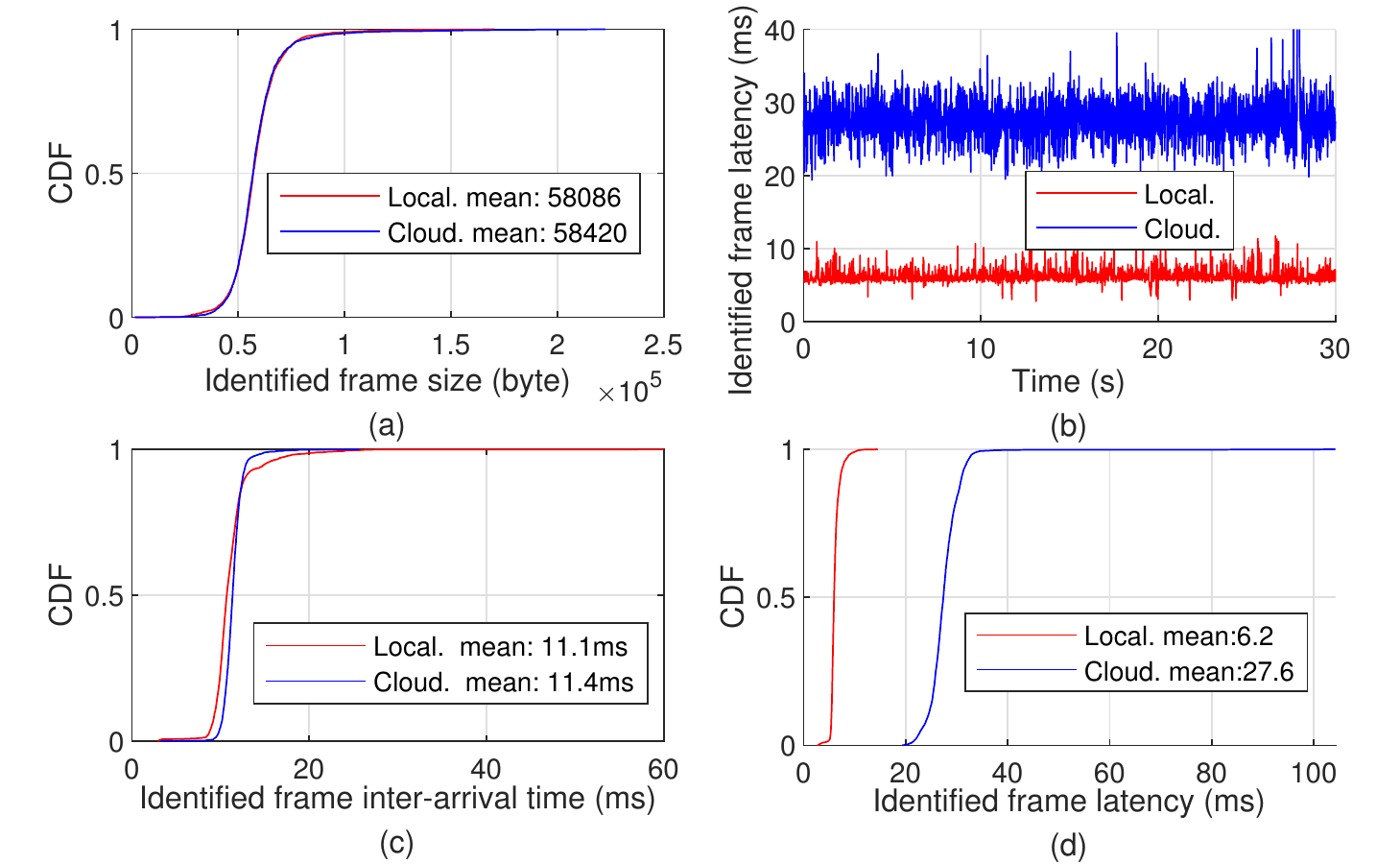}
	\vspace{-0.8cm}
	\caption{Identified video frame metrics for local and cloud cases (Beat Saber). (a) CDF of identified frame size. (b) Identified frame latency vs. time. (c) CDF of identified frame inter-arrival time. (d) CDF of identified frame latency.}
	\label{fig:BeatsaberCDF}
	\vspace{-0.3cm}
\end{figure}

We use the method presented in Section \ref{fsitmethod} to identify video frame size and inter-arrival time. The cumulative distribution functions (CDFs) for the frame size and inter-arrival time in both local and cloud cases are shown in Fig. \ref{fig:BeatsaberCDF} (a) and (c). We find that the identified frame sizes have the same statistical properties for both cloud and local cases. The frame inter-arrival time of the local case is consistent with the 90 Hz setting (11.1 ms frame interval). This result verifies the correctness of our method to identify video frames. The data from the other game, Steam VR Home shows similar patterns. The identified frame size, data rate and inter-arrival time are also shown in the first three ``Normal'' rows in Table \ref{table_allmetrics}, respectively. The data rates are lower than the 40 Mbps streaming bit rate setting. This may be caused by the different internal frame generation mechanisms within different games.


Using the method in Section \ref{lossmethod}, we compute the frame loss rates of the two games. The results are shown in the ``Frame loss rate - Normal'' row in Table \ref{table_allmetrics}. We can see that the local case has very small frame loss rate, indicating a stable network. The loss rate for the cloud case is slightly larger, but is still a small number. It does not affect the quality of gaming experience.

The video frame latency results from the method in Section \ref{fptmethod} are illustrated in Fig. \ref{fig:BeatsaberCDF} (b) and (d). The cloud gaming traffic data has on average 21.4 ms larger latency than the local gaming does. Results for both games in both local and cloud cases are listed in the ``Avg. frame latency - Normal'' row in Table \ref{table_allmetrics}.  This demonstrates a greater latency in cloud VR video streaming as expected. 


%


\subsection{Metrics with Data Rate Limit}
We apply different data rate throttling by setting the WiFi router's data rate limit to 54, 40.5, and 27 Mbps. At the same time, we keep the streaming bit rate setting of Virtual Desktop as fixed to 40 Mbps. We use the same method to identify the video frame size, frame inter-arrival time, frame loss rate, and frame latency as summarized in Table \ref{table_allmetrics}.

\begin{figure}
	\centering
	\includegraphics[width=1\linewidth]{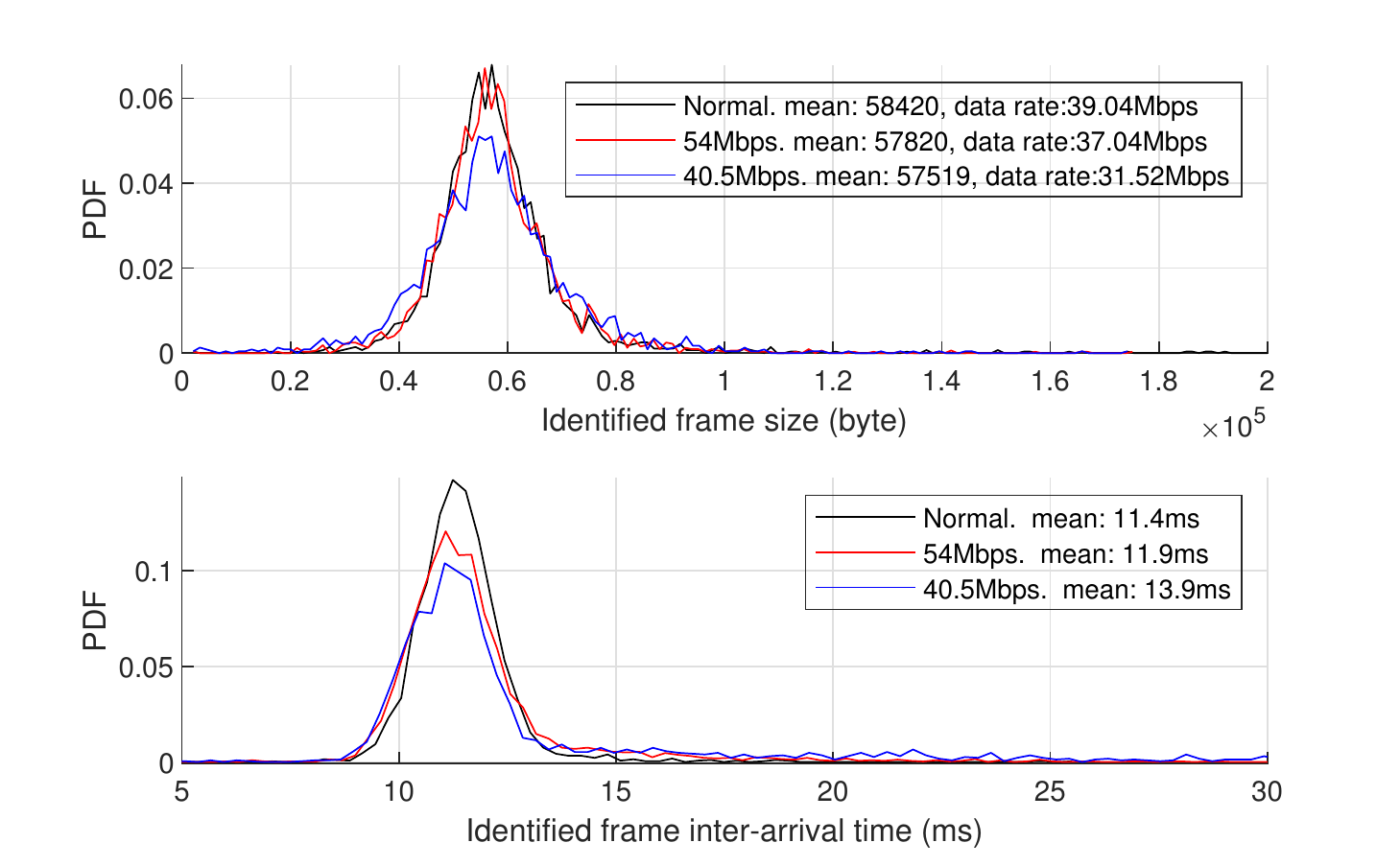}
	\vspace{-0.8cm}
	\caption{PDF of Identified video frame size and inter-arrival time for cloud streaming under data rate limit (Beat Saber, fixed video bit rate setting).}
	\label{fig:BeatSaberthrottlecloudFixPDF}
	\vspace{-0.4cm}
\end{figure}

The probability density function (PDF) curves of the identified video frame size and the frame inter-arrival time for the cloud streaming case with fixed video bit rate setting are shown in Fig. \ref{fig:BeatSaberthrottlecloudFixPDF}. We can see that with a stricter data rate limit, the actual streaming data rate is decreasing. When the throttling is down to 27 Mbps, the video packets from various frames are queued and packets are dropped. In this setup the delay is very high and the VR service is not usable.
The two tested games under both cloud and local case show similar patterns in frame size and inter-arrival time.

\begin{figure}
	\centering
	\includegraphics[width=1\linewidth]{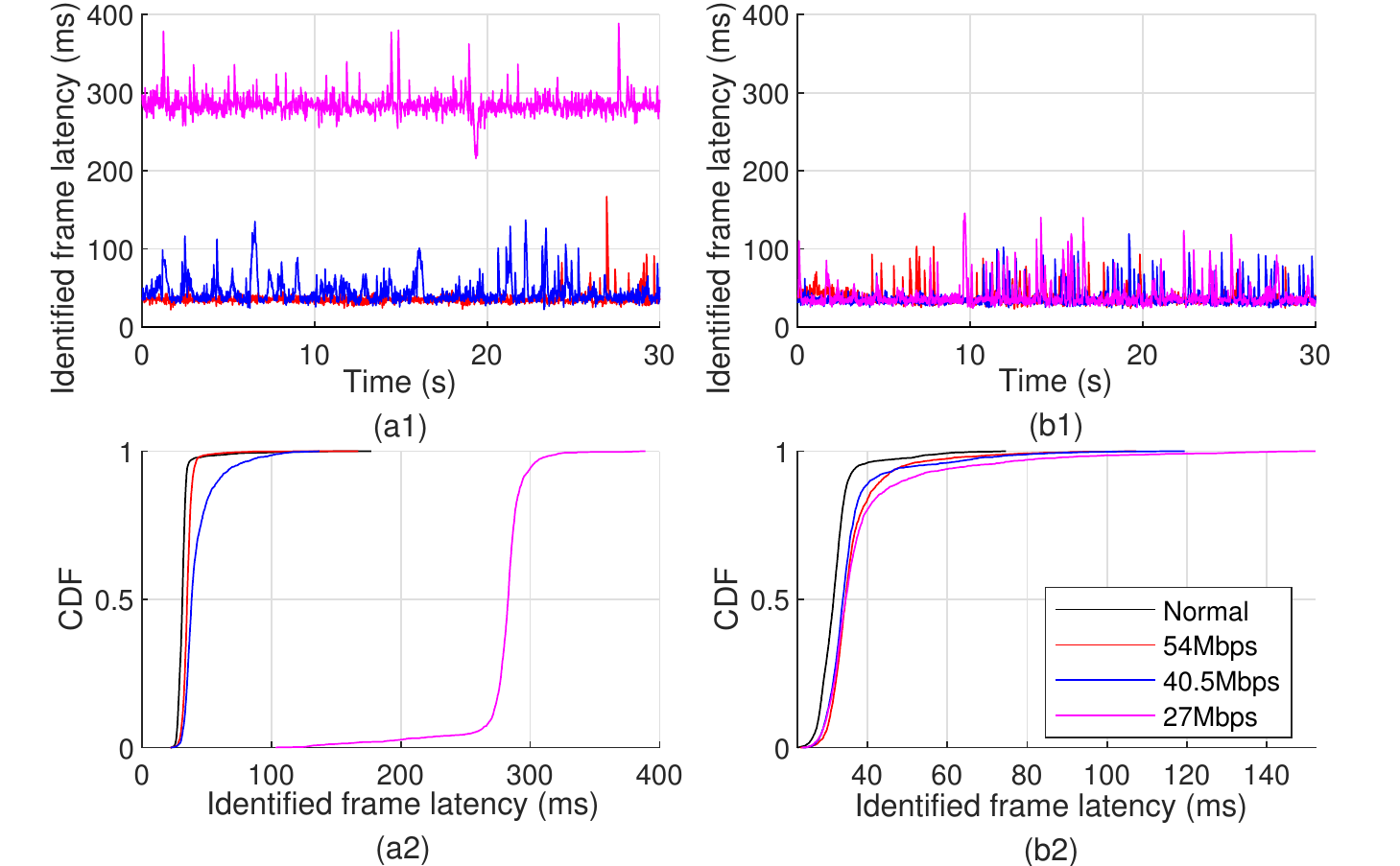}
	\vspace{-0.8cm}
	\caption{Identified video frame latency under different data rate throttling (Steam VR Home, cloud streaming). (a1) and (a2): Fixed video bit rate setting, (b1) and (b2): Adaptive video bit rate setting. ``Normal'' stands for no throttling.
}
	\label{fig:VRhomethrottlecloudfpt}
	\vspace{-0.4cm}
\end{figure}

The frame latency results of Steam VR Home in the cloud streaming case without a data rate limit are shown in Fig. \ref{fig:VRhomethrottlecloudfpt} (a1) and (a2). We can see that the video frame latency and frame loss rate (the last column in Table \ref{table_allmetrics}) increase when we apply tighter data rate throttling. With 40.5 Mbps limit, the actual streaming data rate drops considerably below the 40 Mbps setting, causing discontinuity in video frames and perceivable latency. When the data rate limit is 27 Mbps, the identified frame loss rate is up to 32.51\% and the latency is around 280 ms, which causes an unacceptable VR gaming experience.


\begin{table}[!t]
	\begin{threeparttable}
	\vspace{-0.1cm}
		\caption{VR Traffic Characteristics with Fixed Video Bit Rate}
		\label{table_allmetrics}
		\centering
		\begin{tabular}{p{1.5cm} c r r r r}
			\toprule
			\multirow{2}{1.5cm}{\centering{Metrics}}&\multirow{2}{1.6cm}{\centering{Data Rate Limit (Mbps)}}
			&\multicolumn{2}{c}{Beat Saber}&\multicolumn{2}{c}{Steam VR Home}\\
			\cline{3-6}
			\multirow{2}{*}{} & &Local& Cloud& Local&Cloud\\
			\hline
			\multirow{4}{1.5cm}{\centering{Avg. frame size (byte)}}
			&Normal& 58087 & 58420& 57957&58300\\
			& 54  & 57613 & 57820& 58933&58394\\
			& 40.5& 58146 & 57519& 57591&58228\\
			& 27  & 49228 & 49602& 51567& 43737\\
			\cline{2-6}
			\multirow{4}{1.5cm}{\centering{Data rate (Mbps)}} 
			& Normal&39.76 & 39.04& 34.37& 33.46\\
			& 54   &38.38 & 37.04& 33.30& 29.26\\
			& 40.5&32.66 & 31.52& 30.48& 29.70\\
			& 27 & 21.83 & 21.85& 21.80& 21.87\\
			\cline{2-6}
			\multirow{4}{1.5cm}{\centering{Avg. frame inter-arrival time (ms)}}
			& Normal& 11.1 & 11.4& 12.9& 13.3\\
			& 54   & 11.5 & 11.9& 13.5& 15.2\\
			& 40.5 & 13.6 & 13.9& 14.6& 15.0\\
			& 27   & 17.2 & 17.3& 18.0& 15.3\\
			\cline{2-6}
			\multirow{4}{1.5cm}{\centering{Frame loss rate (\%)}} 
			& Normal& 0.04 & 0.75& 0.04& 0.75\\
			& 54  & 0.03 & 5.06& 0.04& 1.29\\
			& 40.5 & 0.20 & 14.96& 0.09& 10.56\\
			& 27   & 0.72 & 28.30& 1.78& 32.51\\
			\cline{2-6}
			\multirow{4}{1.5cm}{\centering{Avg. frame latency (ms)}}
			& Normal & 6.2 & 27.6& 5.9& 31.9\\
			& 54 & 18.2 & 36.4& 14.3& 35.4\\
			& 40.5 & 55.3 & 63.3& 29.2& 43.1\\
			& 27 & 293.9 & 280.0& 287.9& 279.4\\
			\bottomrule
		\end{tabular}
		
		\begin{tablenotes}[para,flushleft]
			Note: The metrics are computed by the video frame identification algorithm. ``Normal'' represent no limitation on the data rate.
		\end{tablenotes}
	\end{threeparttable}
	\vspace{-0.4cm}
\end{table}



\section{VR Video Frame Identification and Modeling with Adaptive Bit Rate}
\subsection{Identified Metrics}
We change the setting of Virtual Desktop to ``Adaptive bit rate'' with the maximum unchanged at 40 Mbps. We collect data in different throttling conditions for both local and cloud gaming schemes. The video frame latency and frame loss rate of Steam VR home in the cloud case are shown in Fig. \ref{fig:VRhomethrottlecloudfpt} (b1) and (b2). We can see that compared with the fixed bit rate setting, the frame latency does not change much with varying data rate limit. Especially for 27 Mbps throttling, the frame latency does not surge as much as in the fixed video bit rate case. All the computed metrics are listed in Table \ref{table_allmetricsADP}.

As can be seen from Table \ref{table_allmetricsADP}, similar to the fixed video bit rate case in Section \ref{fixedrate}, the frame size and data rate decreases with reducing data rate limit. The inter-arrival time remains stable for all cases, showing the adaptive adjustment on the frame size according to the available data rate. The frame loss rate remains stable in the local case for both games, and is much smaller than the fixed bit rate case for both games in both the local and cloud schemes as shown in Table \ref{table_allmetrics}. In the cloud case for both games, the frame loss is larger than that in the local case, and is increasing with smaller data rate limit. The frame latency also remains a small number even with low data rate. This results in a smooth game play with lower image quality even in the stringent 27 Mbps throttling case. From the player's perspective, the game is running smoothly with unnoticeable latency and frame loss although the graphic quality is lowered.

\begin{table}[!t]
	\begin{threeparttable}
	\vspace{-0.1cm}
		\caption{VR Traffic Characteristics with Adaptive Video Bit Rate}
		\label{table_allmetricsADP}
		\centering
		\begin{tabular}{p{1.5cm} c r r r r}
			\toprule
			\multirow{2}{1.5cm}{\centering{Metrics}}&\multirow{2}{1.6cm}{\centering{Data Rate Limit (Mbps)}}
			&\multicolumn{2}{c}{Beat Saber}&\multicolumn{2}{c}{Steam VR Home}\\
			\cline{3-6}
			\multirow{2}{*}{} & &Local& Cloud& Local&Cloud\\
			\hline
			\multirow{4}{1.5cm}{\centering{Avg. frame size (byte)}}
			&Normal& 57376 & 57719& 57896&57943\\
			& 54  & 53052 & 46976& 52403&48771\\
			& 40.5& 40668 & 39093& 40888&38985\\
			& 27  & 29103 & 24808& 29488& 28468\\
			\cline{2-6}
			\multirow{4}{1.5cm}{\centering{Data rate (Mbps)}} 
			& Normal&39.46 & 39.58& 33.90& 32.62\\
			& 54   &36.09 & 31.55& 29.53& 28.40\\
			& 40.5&27.32 & 26.10& 22.85& 21.65\\
			& 27 & 19.13 & 18.46& 16.00& 15.83\\
			\cline{2-6}
			\multirow{4}{1.5cm}{\centering{Avg. frame inter-arrival time (ms)}}
			& Normal& 11.1 & 11.1& 13.0& 13.6\\
			& 54   & 11.2 & 11.4& 13.5& 13.1\\
			& 40.5 & 11.4 & 11.4& 13.7& 13.7\\
			& 27   & 11.6 & 11.8& 14.1& 13.7\\
			\cline{2-6}
			\multirow{4}{1.5cm}{\centering{Frame loss rate (\%)}} 
			& Normal& 0.03 & 1.56& 0.04& 1.06\\
			& 54  & 0.07 & 1.91& 0.04& 3.05\\
			& 40.5 & 0.03 & 2.24& 0.04& 3.53\\
			& 27   & 0.07 & 5.12& 0.04& 4.78\\
			\cline{2-6}
			\multirow{4}{1.5cm}{\centering{Avg. frame latency  (ms)}}
			& Normal & 6.8 & 31.9& 6.1& 32.0\\
			& 54 & 14.7 & 32.3& 13.2& 36.4\\
			& 40.5 & 16.9 & 34.1& 13.7& 35.8\\
			& 27 & 23.1 & 37.5& 15.3& 38.5\\
			\bottomrule
		\end{tabular}
		
	\end{threeparttable}
	\vspace{-0.4cm}
\end{table}

\subsection{Statistical Modeling}
We present the statistical models for the identified frame size and inter-arrival time. Since in practice, there are limitations in security and privacy, it is important to model the traffic using limited features \cite{navarro2020survey}. With these models, researchers are able to generate the VR video traffic by simulation.

\begin{figure}
	\centering
	\vspace{-0.2cm}
	\includegraphics[width=1\linewidth]{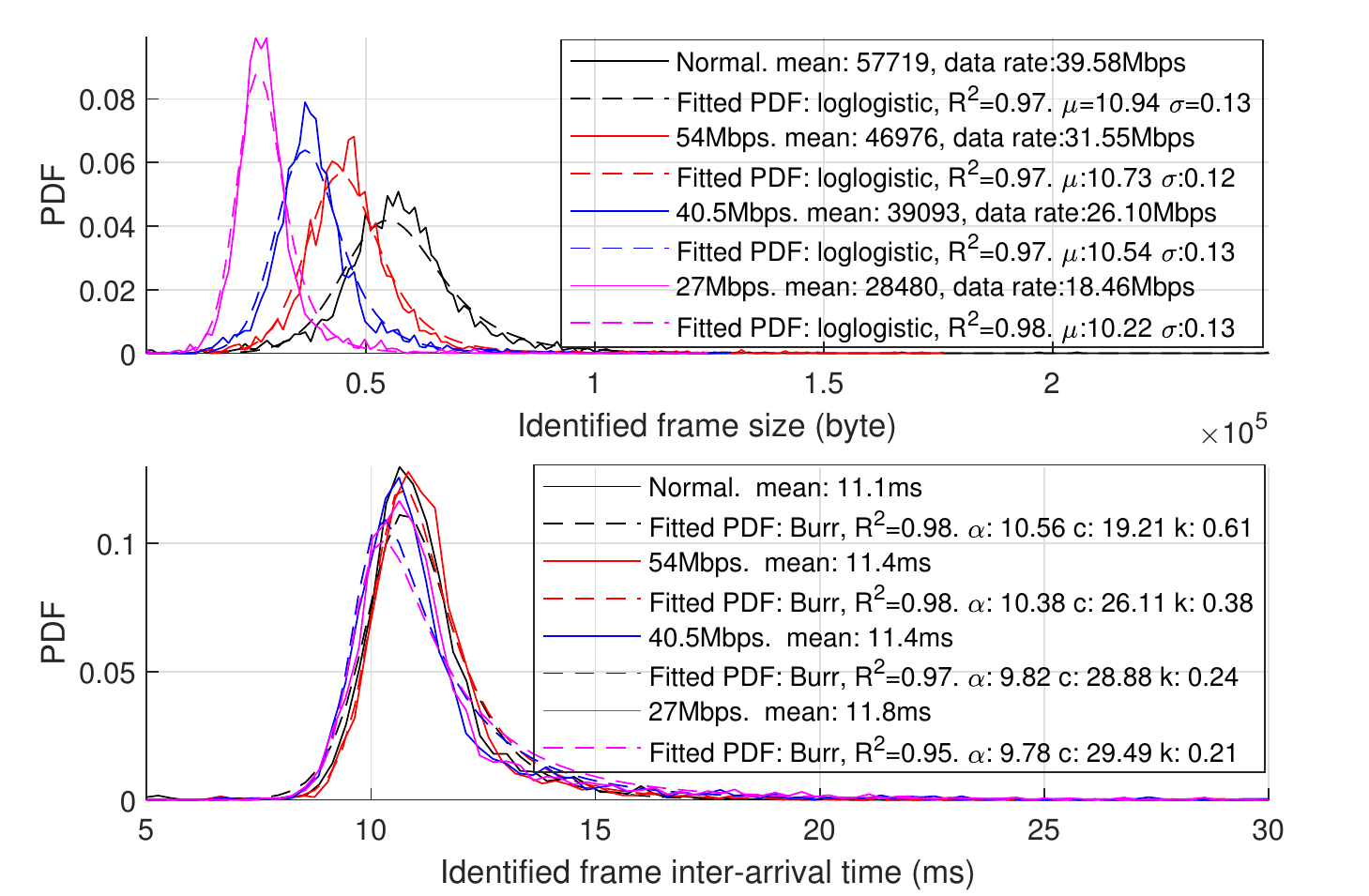}
	\vspace{-0.8cm}
	\caption{Statistical modeling for frame size and inter-arrival time (Beat Saber, cloud streaming with adaptive video bit rate). ``loglogistic'' and ``Burr'' in the legend are names of the fitted distributions, followed by $R^2$ and corresponding parameters.}
	\label{fig:CloudAdpFramesizeintertBeatSaber}
	\vspace{-0.3cm}
\end{figure}

By observing the PDF curves shown by the solid lines in Fig \ref{fig:CloudAdpFramesizeintertBeatSaber}, we can see that both the frame size and inter-arrival time have asymmetric distributions with heavier tails on the right side, which is similar to the feature of loglogistic and Burr distributions.

We model the video frame size as loglogistic distribution, which has the general form of PDF written as
\begin{align}
	\text{loglogistic: } f(x)=\frac{\exp{(\frac{\ln{x}-\mu}{\sigma})}}{\sigma x \left(1+\exp{(\frac{\ln{x}-\mu}{\sigma})}\right)^2}, \; x\geq 0,
\end{align}
where parameter $\mu$ is the mean of the logarithmic values and $\sigma$ is the scale parameter \cite{tadikamalla1980look}. The hump of the PDF curve will be lower and wider in the $+x$ direction with a larger $\mu$. A larger $\sigma$ will reduce the height of the hump and make both the left and right tails heavier.

We use the Burr distribution to model the video frame inter-arrival time. The PDF of Burr distribution is
\begin{align}
	\text{Burr: } f(x)=\frac{\frac{kc}{\alpha}\left(\frac{x}{\alpha}\right)^{c-1}}{\left(1+\left(\frac{x}{\alpha}\right)^c\right)^{k+1}}, \; x\geq 0,
\end{align}
where $c$ and $k$ are the shape parameters and $\alpha$ is the scale parameter \cite{tadikamalla1980look}. The peak of the curve becomes higher and the right tail becomes heavier when $c$ and $k$ increase. Given a larger $\alpha$, the peak will be shorter and the curve will move toward $+x$.

The computed distribution parameters for frame size and inter-arrival time with different data rate limits in both local and cloud cases are listed in Table \ref{table_allmodels}. As an example, we show in Fig. \ref{fig:CloudAdpFramesizeintertBeatSaber} the fitted PDF curves and the corresponding identified parameters of frame size and inter-arrival time for Beat Saber in the cloud case with adaptive video bit rate. The coefficient of determination $R^2$ \cite{nagelkerke1991note} is also larger than 0.95 in the figure. We can see that the PDF curves of the loglogistic and Burr distributions fit the data well, demonstrating that the statistical models represent the real traffic behavior. 

\begin{table}[!t]
	\begin{threeparttable}
		\caption{Statistical Model Parameters of VR Video Traffic}
		\label{table_allmodels}
		\centering
		\begin{tabular}{p{1.1cm} c r r r r r}
			\toprule
			\multirow{2}{1.1cm}{\centering{Game}}&\multirow{2}{1.8cm}{\centering{Data Rate Limit (Mbps)*}}
			&\multicolumn{2}{c}{Frame Size }&\multicolumn{3}{c}{Inter-arrival Time }\\
			\cline{3-7}
			\multirow{2}{*}{} & &$\mu$& $\sigma$& $\alpha$& $c$ & $k$\\
			\hline
			\multirow{8}{1.1cm}{\centering{Beat Saber}}
			&Normal/L& 10.95 & 0.11& 9.72&21.29 &0.33\\
			& 54/L  & 10.86 & 0.10& 9.41&21.91&0.26\\
			& 40.5/L& 10.59 & 0.12& 9.23&25.22&0.20\\
			& 27/L  & 10.24 & 0.14& 9.12& 23.97&0.19\\
			\cline{2-7}
			&Normal/C& 10.94& 0.13& 10.56&19.21&0.61\\
			& 54/C  & 10.73& 0.12& 10.38&26.11&0.38\\
			& 40.5/C& 10.54& 0.13& 9.82&28.88&0.24\\
			& 27/C  & 10.22& 0.13& 9.78& 29.49&0.21\\
			\cline{1-7}
			\multirow{8}{1.1cm}{\centering{Steam VR Home}}
			&Normal/L& 10.96 & 0.12& 12.35&39.47 &0.39\\
			& 54/L  & 10.83 & 0.17& 12.97&12.49&0.79\\
			& 40.5M/L& 10.58 & 0.16& 12.98&26.14&0.55\\
			& 27/L  & 10.25 & 0.16& 13.17& 33.48&0.39\\
			\cline{2-7}
			&Normal/C& 10.91 & 0.19& 13.27&24.91 &0.76\\
			& 54/C  & 10.74 & 0.20& 12.10&46.94&0.26\\
			& 40.5/C& 10.47 & 0.25& 12.82&27.87&0.44\\
			& 27/C  & 10.91 & 0.19& 12.34& 30.96&0.30\\
			\bottomrule
		\end{tabular}
		
		\begin{tablenotes}[para,flushleft]
			Note: * ``/L'' and ``/C'' represent the local and cloud streaming cases,  respectively. The frame size and the inter-arrival time are modeled as loglogistic ($\mu$ and $\sigma$) and Burr ($\alpha$, $c$ and $k$) distributions, respectively. 
		\end{tablenotes}
	\end{threeparttable}
	\vspace{-0.4cm}
\end{table}

%
%
%

\section{Conclusion and Future Work}
Cloud gaming is becoming a promising technique, which enables users to play high quality VR games without equipping high end gaming PCs. We take a very first step to understand the cloud VR gaming traffic data using a real commercial cloud gaming server. We implement a testbed for traffic data collection based on the Paperspace cloud server, the  Oculus Quest 2 VR headset, and an off-the-shelf WiFi router. We analyze the cloud gaming traffic data characteristics for video, audio and control data flows from different games, and compare them with the local streaming scheme. We develop a method to identify the application-layer video frames using only the information from raw packets without decoding or packet inspection. With the frame identification result, we evaluate the traffic behaviors and find that the adaptive bit rate enables shorter latency and less frame loss and thus a better gaming experience than the fixed bit rate. We show that the video frame size and inter-arrival time can be modeled as loglogistic and Burr distributions, respectively, which are helpful in VR traffic simulation and are useful in real-world applications such as RRM and network optimization. Our VR traffic data-sets are available at \cite{VRdataset}.

We find that varying latency and data rate limit are challenging factors for the quality of cloud VR gaming services. Therefore, future cloud VR gaming servers need to have a more resilient design to reduce the latency and combat the instability of the Internet, so as to offer better visual quality and immersive experience to VR players.

In the future, we will investigate the relation among the traffic characteristics such as the joint distribution of the frame size and inter-arrival time, study more cases with diverse setting combinations, such as more different games, adaptive throttling, and varying user motion, and validate the frame size and inter-arrival time models that we developed in these cases.

\bibliographystyle{IEEEtran}
\bibliography{IEEEabrv,paper}

\end{document}